\documentclass[aps,prb,preprint,superscriptaddress]{revtex4-1}
\usepackage{graphicx}
\usepackage{dcolumn}
\usepackage{bm}
\usepackage{epstopdf}
\usepackage{hyperref}
\usepackage[utf8]{inputenc}
\usepackage[T1]{fontenc}
\usepackage{lmodern}
\usepackage{amsmath}
\usepackage{amssymb}
\usepackage{booktabs}
\usepackage{color}
\usepackage{float}
\usepackage[table]{xcolor}
\definecolor{Gray}{gray}{0.9}
\newcolumntype{m}{>{\columncolor{Gray}}c}

\begin{document}


\title{Magnetic phase diagram of the quantum spin chain compound SrCo$_{2}$V$_{2}$O$_{8}$: a single-crystal neutron diffraction study}


\author{L. Shen}
\affiliation{School of Physics and Astronomy, University of Birmingham, Birmingham B15 2TT, United Kingdom}
\affiliation{Division of Synchrotron Radiation Research, Lund University, SE-22100 Lund, Sweden}
\affiliation{Stanford Institute for Materials and Energy Science, Stanford University and SLAC National Accelerator Laboratory, Menlo Park, California 94025, USA}
\affiliation{Linac Coherent Light Source, SLAC National Accelerator Laboratory, 2575 Sand Hill Road, Menlo Park, CA 94025, USA}%

\author{O. Zaharko}
\affiliation{Laboratory for Neutron Scattering and Imaging, Paul Scherrer Institut, CH-5232 Villigen PSI, Switzerland}

\author{J. O. Birk}
\affiliation{Laboratory for Neutron Scattering and Imaging, Paul Scherrer Institut, CH-5232 Villigen PSI, Switzerland}

\author{E. Jellyman}
\affiliation{School of Physics and Astronomy, University of Birmingham, Birmingham B15 2TT, United Kingdom}

\author{Z. He}
\affiliation{State Key Laboratory of Structural Chemistry, Fujian Institute of Research on the Structure of Matter, Chinese Academy of Sciences, Fuzhou, Fujian 350002, China}

\author{E. Blackburn}
\affiliation{School of Physics and Astronomy, University of Birmingham, Birmingham B15 2TT, United Kingdom}
\affiliation{Division of Synchrotron Radiation Research, Lund University, SE-22100 Lund, Sweden}



\begin{abstract}
We explore magnetic order in the quantum spin chain compound SrCo$_{2}$V$_{2}$O$_{8}$ up to 14.9\,T and down to 50\,mK, using single-crystal neutron diffraction. Upon cooling in zero-field, commensurate antiferromagnetic (C-AFM) order with modulation vector \textit{\textbf{k}}$_\mathrm{{C}}$ = (0, 0, 1) develops below $T_\mathrm{{N}}$\ $\simeq$\ 5.0\ K. Applying an external magnetic field (\textit{H}\ $\parallel$\ \textit{c} axis) destabilizes this C-AFM order, leading to an order-disorder transition between $T_\mathrm{{N}}$ and $\sim$ 1.5\,K. Below 1.5\,K, a commensurate to incommensurate (IC-AFM) transition occurs at 3.9\,T, above which the magnetic reflections can be indexed by \textit{\textbf{k}}$_{IC}$ = (0, 0, 1\,$\pm$\,$\delta{l}$). The incommensurability $\delta{l}$ scales monotonically with \textit{H} until the IC-AFM order disappears around 7.0\,T. Magnetic reflections modulated by \textit{\textbf{k}}$_\mathrm{{C}}$ emerge again at higher fields. While the characters of the C-AFM, IC-AFM and the emergent AFM order in SrCo$_{2}$V$_{2}$O$_{8}$ appear to fit the descriptions of the N\'eel, longitudinal spin density wave and transverse AFM order observed in the related compound BaCo$_{2}$V$_{2}$O$_{8}$, our results also reveal several unique signatures that are not present in the latter, highlighting the inadequacy of mean-field theory in addressing the complex magnetic order in systems of this class.
\end{abstract}

\maketitle
\section{Introduction}\label{Intro}
Magnetic field is a very important parameter when tuning the physical properties in quasi one-dimensional (1D) spin-1/2 magnets. The magnetic excitation spectrum of a single quantum chain is a continuum composed of pairs of spinons, each with \textit{S}\ =\ 1/2, that can propagate like domain walls\ \cite{Faddeev}. In quasi 1D magnets where there are non-zero interactions between the chains, the spinons become confined by an attractive potential\ \cite{Lake3}. Concomitantly, the continuum spectrum is replaced by a series of discrete spinon bound states. It has been found that the spinon confinement can be significantly tuned by applying a magnetic field\ \cite{Kenzelmann,Coldea,Wang,Faure,Wang2}.

Furthermore, exotic magnetic long-range order (LRO) may appear in a magnetic field. For example, in weakly coupled spin chains or ladders with a singlet-dimer ground state (\textit{S}\ =\ 0), applying a magnetic field splits the associated triplet excitation (\textit{S}\ =\ 0,\ $\pm$\,1); a singlet-dimer to LRO transition occurs at the closure of the energy gap corresponding to the lowest triplet branch (\textit{S}\ =\ 1). This transition, also known as magnon Bose-Einstein condensation (BEC), has been intensively studied in the last two decades\ \cite{Ruegg, Giamarchi, Zapf}.

Recently, the weakly coupled quantum spin chain compound SrCo$_{2}$V$_{2}$O$_{8}$ has raised much attention due to the exotic magnetism that it hosts, including the magnetic-field-induced order-disorder transition \cite{He2}, spinon confinement\ \cite{Wang,Bera3} and Bethe strings\ \cite{Wang2}. This compound crystallizes in a body-centered tetragonal lattice (space group $I4_\mathrm{{1}}cd$), in which 4-fold screw chains of CoO$_{6}$-octahedra run along the crystallographic \textit{c}-axis\ \cite{He,Bera}. These chains are well separated in the \textit{ab} plane, greatly reducing the strength of the interactions between the chains.\ The Co$^{2+}$ ion (3$d^\mathrm{{7}}$) has an effective spin of 1/2 because of the octahedral distortion and spin-orbit coupling\ \cite{Abragam, Lines}. The intrachain spin interactions in SrCo$_{2}$V$_{2}$O$_{8}$ can be described by an XXZ model written as
\begin{equation}\label{Ham}
\mathcal{H}_{XXZ} = J \sum_{i} \big\{S^{z}_{i}S^{z}_{i+1} + \epsilon
(S^{x}_{i}S^{x}_{i+1} + S^{y}_{i}S^{y}_{i+1})\big\} - g_{z}\mu_{B}\sum_{i}S^{z}_{i}H,
\end{equation}
where \textit{J}\ >\ 0 is the nearest-neighbour (NN) antiferromagnetic (AFM) exchange constant, $\epsilon$ is the anisotropy parameter, and $g_{z}$ is the component of the Land\'e \textit{g}-tensor along the chain direction\ \cite{Bera3,Wang,Wang2}. This Hamiltonian can be exactly solved by the Bethe ansatz\ \cite{Bethe}. The anisotropy in SrCo$_{2}$V$_{2}$O$_{8}$ is Ising-like ($\epsilon$\ <\ 1)\ \cite{He,Wang,Wang2,Bera,Bera3}, for which the XXZ model predicts a N\'eel type AFM ground state. A spin disordered state, described as a Tomonaga-Luttinger liquid (TLL), can be induced above a critical magnetic field. The spinon spectrum in the TLL state of an Ising-like XXZ spin chain is dominated by a longitudinal (transverse) mode in the intermediate (high) magnetic field region\ \cite{Haldane2,Bogoliubov}. 
In a quasi 1D system, the interchain exchange interactions come into play at low temperatures. Consequently, the longitudinal (transverse) mode is expected to condense, giving rise to a longitudinal spin density wave (transverse AFM) state. In $M$Co$_{2}$V$_{2}$O$_{8}$ ($M$ = Sr, Ba), this spin density wave should be characterized by an incommensurate modulation vector ($\delta$\textit{l}),
\begin{equation}\label{kz}
{|}\delta{}{l}{|} = 4\ \times\ <S_\mathrm{{z}}>\ =\ 4\ \times\ M_\mathrm{{z}}\ /\ g_{z}\mu_\mathrm{{B}},
\end{equation}
where $M_\mathrm{{z}}$ is the uniform magnetization along the chain direction \cite{Kimura2,Canevet}.

In SrCo$_{2}$V$_{2}$O$_{8}$, the interchain exchange interactions are nonnegligible, leading to the occurrence of N\'eel order at a finite temperature $T_\mathrm{{N}}$\ =\ 5.0\,K\ \cite{He,Bera}. Theoretically, Okunishi and Suzuki have used bosonization combined with a mean-field treatment of the interchain interactions to study the magnetic phase diagram in $M$Co$_{2}$V$_{2}$O$_{8}$ ($M$ = Sr, Ba) \cite{Okunishi}. At very low temperatures, they obtained a longitudinal spin density wave (LSDW) phase that replaces the N\'eel order above an intermediate critical field, and another LSDW to transverse AFM phase transition in the high field region. So far, the exotic magnetic-field-induced phases have only been confirmed in the Ba-compound by single-crystal neutron diffraction\ \cite{Kimura,Kimura2,Canevet,Grenier2}. These studies have revealed that the critical field for the N\'eel to LSDW transition ($\sim$ 3.9\,T) agrees well with the mean-field prediction, while the one for the LSDW to transverse AFM transition ($\sim$ 9.0 T) appears to be much lower than the predicted value ($\sim$ 15.1 T). The discrepancy might indicate that the inter- and intra- chain exchange parameters used in Ref.~\onlinecite{Okunishi} are not correct\ \cite{Grenier2} or that the mean-field theory is inadequate. As pointed out by a recent nuclear magnetic resonance (NMR) study, the intra- and inter- couplings in BaCo$_{2}$V$_{2}$O$_{8}$ are rather complicated and could be strongly field-dependent\ \cite{Klanjsek2}.

The static magnetic order in SrCo$_{2}$V$_{2}$O$_{8}$ has only been investigated in zero-field by neutron powder diffraction \cite{Bera}, while its evolution in a longitudinal magnetic field, especially in the low temperature region where the LSDW and transverse AFM are predicted to develop \cite{Okunishi}, is not known. To further check the mean-field theory, and more importantly, obtain more insights into the magnetic properties in materials of this class, we report the magnetic phase diagram of SrCo$_{2}$V$_{2}$O$_{8}$ up to 14.9\,T ($H\,\parallel\,c$ axis) and down to 50 mK, using single-crystal neutron diffraction. Our experimental results are organized into two sections. The first focuses on the commensurate (C) AFM order and magnetic-field-induced order-disorder transition above 2.0 K (Section\ \ref{HT}). The second looks at the phase diagram below 2.0 K for fields up to 14.9 T, covering a magnetic-field-induced incommensurate (IC) AFM order (3.9 T < $\mu_\mathrm{0}H$ < 7.0 T) and emergent C-AFM order ($\mu_\mathrm{0}H$ > 7.0 T) (Section\ \ref{LT}). Finally, in Section\ \ref{Discussion}, we will discuss several unique signatures of the spin states in SrCo$_{2}$V$_{2}$O$_{8}$ that are not present in BaCo$_{2}$V$_{2}$O$_{8}$.

\section{Experimental methods}
Two high quality SrCo$_{2}$V$_{2}$O$_{8}$ single crystals ($\sim$\ 3\,$\times$\,3\,\,$\times$\,6\ mm$^{3}$) were measured in this investigation. They were grown by the spontaneous nucleation method described in Ref.~\onlinecite{He}. All single-crystal neutron diffraction measurements were carried out at the Swiss Spallation Neutron Source (SINQ) at the Paul Scherrer Institute. Both single crystals were aligned using the two-axis neutron diffractometer ORION. The first single-crystal was mounted into a dilution refrigerator and then into a 6\,T vertical cryomagnet, with the magnetic field applied along the \textit{c}-axis.\ This was installed on the thermal neutron diffractometer TriCS and measured with a lifted detector (normal beam geometry). For the magnetic structure determination, a set of magnetic and nuclear reflections were collected at neutron wavelength $\lambda$\ =\ 1.178\,\AA{} using a Ge\,(311) monochromator. All the other measurements were performed at $\lambda$\ =\ 2.317\,\AA{} using a PG\,(002) monochromator with a vertical 80$^{\prime}$ collimator installed to improve the resolution along the $\textbf{c}^{\star}$ direction in the reciprocal space.

The second crystal was mounted in a 15\,T vertical cryomagnet equipped with a dilution refrigerator insert and then measured on the cold triple-axis spectrometer RITA-II. The \textit{c}-axis was aligned along the magnetic field. The incident neutron energy was fixed at 5\,meV using a vertically focusing PG\,(002) monochromator. The energy of the scattered neutrons was analysed using a multi-blade PG\,(002) crystal analyzer, which was operated in a monochromatic imaging mode\,\cite{rita}. A cooled beryllium filter was placed between the sample and analyzers to suppress the $\lambda$/2 contamination. The neutrons were detected using a position-sensitive detector (PSD) consisting of 128\,$\times$\,128 pixels. In all measurements, only the elastic scattering signal was recorded.

\section{N\'eel order and magnetic-field-induced order-disorder transition above 2.0 K}\label{HT}

\begin{figure}[H]
	\centering
	\includegraphics[width=0.96\textwidth]{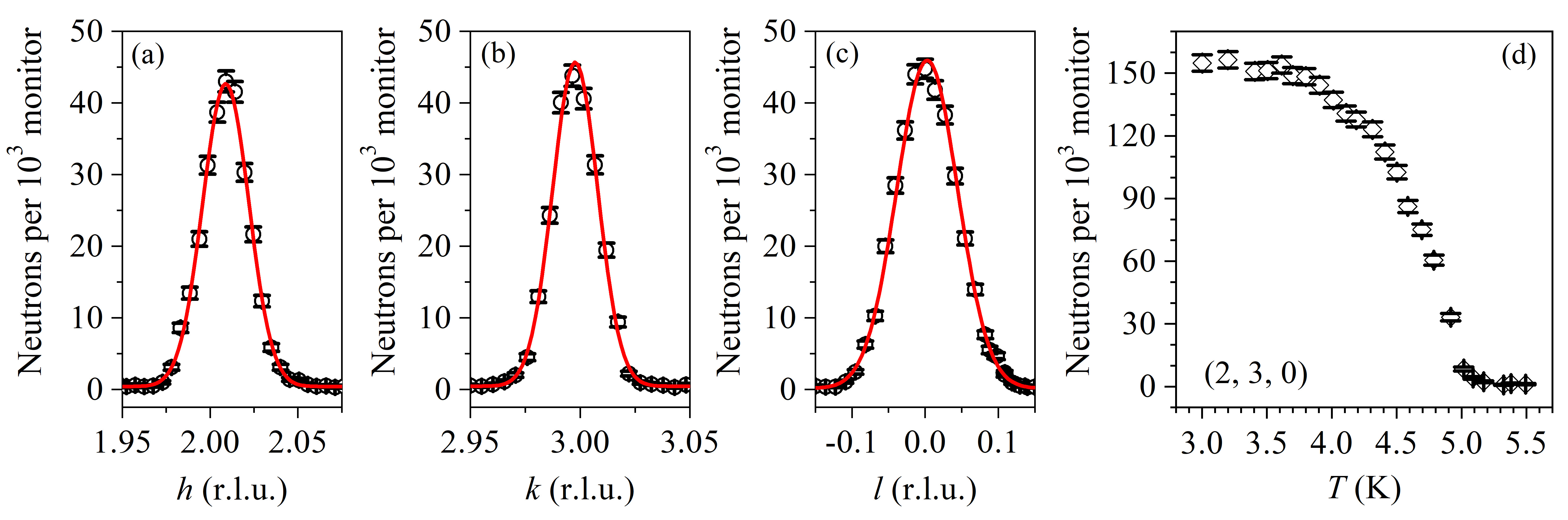}
	\caption{(a-c) Open symbols: \textit{hkl}-scans performed at \textit{T}\ =\ 50\,mK and $\mu_\mathrm{{0}}H$ \ =\ 0\,T. The red lines are numerical fits using a Gaussian function. (d) Temperature dependence of the peak intensity of the (2, 3, 0) reflection at $\mu_\mathrm{{0}}H$ \ =\ 0\,T.}	
	\label{fig:1}
\end{figure}

Several crystallographically forbidden reflections are observed at \textit{T}\ =\ 50\,mK and $\mu_\mathrm{{0}}H$ \ =\ 0\,T. They can be indexed by the commensurate modulation vector $k_\mathrm{{C}}$ = (0, 0, 1), which is consistent with the N\'eel order reported in the previous neutron powder diffraction study\ \cite{Bera}. Fig.\,\ref{fig:1}a-c shows cuts through the (2, 3, 0) reflection, measured on TriCS. Each peak was fitted using a Gaussian function, giving full width at half maximum (FWHM) values of 0.0310(5), 0.0242(4), and 0.098(1) r.\,l.\,u.\ for the \textit{h}-, \textit{k}-, and \textit{l}- scans, respectively. These values serve as the TriCS instrumental resolution parameters around this particular momentum transfer referred to below. The temperature dependence measurements reveal that this reflection disappears around 5.0 K (Fig.\,\ref{fig:1}d), which agrees with the $T_\mathrm{{N}}$ extracted from a previous heat capacity investigation on the same crystal\ \cite{He}.

To study the magnetic structure in the N\'eel phase, we carried out a representational analysis using the SARAh Representational Analysis software\ \cite{Wills}. Five irreducible representations ($\Gamma_\mathrm{{n}}$, \textit{n}\ =\ 1, 2, 3, 4, 5) could be obtained\ \cite{Bera}. $\Gamma_\mathrm{{1}}$, $\Gamma_\mathrm{{2}}$, $\Gamma_\mathrm{{3}}$, and $\Gamma_\mathrm{{4}}$ did not yield satisfactory agreements to our data.\ This leaves $\Gamma_\mathrm{{5}}$ (Table\,\ref{tab:I}) for our refinement, which is consistent with the conclusion in a neutron powder diffraction study\ \cite{Bera}. For the nuclear structure, we have collected 17 reflections to refine the scale factor, while the atomic positions and isotropic atomic displacement parameters were fixed to the values reported in Ref.~\onlinecite{Bera}.

\begin{table}[H]
  \centering
  \caption{Basis functions $\Psi_\mathrm{{n}}$ (\textit{n}\ =\ 1, ..., 12) of Co for the irreducible representation $\Gamma_\mathrm{{5}}$. The basis functions contributing to a finite moment in the $ab$ plane are highlighted in grey. The atomic sites are labelled as Co$_\mathrm{{1}}$: (\textit{x}, \textit{y}, \textit{z}), Co$_\mathrm{{2}}$: (-\textit{x}\ +\ 1, -\textit{y}\ +\ 1, \textit{z}), Co$_\mathrm{{3}}$: (-\textit{y}\ +\ 1, \textit{x}\ + 1/2, \textit{z}\ +\ 1/4), Co$_\mathrm{{4}}$: (\textit{y}, -\textit{x}\ +\ 1/2, \textit{z}\ +\ 1/4), Co$_\mathrm{{5}}$: (-\textit{x}\ +\ 1, \textit{y}, \textit{z}\ +\ 1/2), Co$_\mathrm{{6}}$: (\textit{x}, -\textit{y}\ +\ 1, \textit{z}\ +\ 1/2), Co$_\mathrm{{7}}$: (\textit{y}\ +\ 1/2, \textit{x}, \textit{z}\ +\ 1/4), and Co$_\mathrm{{8}}$: (-\textit{y}\ +\ 1/2, -\textit{x}\ +\ 1, \textit{z}\ +\ 1/4).}
  \label{tab:I}
  \begin{ruledtabular}
  \begin{tabular}{c|mmcmmcmmcmmc}
Site Label&$\Psi_\mathrm{{1}}$&$\Psi_\mathrm{{2}}$&$\Psi_\mathrm{{3}}$&$\Psi_\mathrm{{4}}$&$\Psi_\mathrm{{5}}$&$\Psi_\mathrm{{6}}$&$\Psi_\mathrm{{7}}$&$\Psi_\mathrm{{8}}$&$\Psi_\mathrm{{9}}$&$\Psi_\mathrm{{10}}$&$\Psi_\mathrm{{11}}$&$\Psi_\mathrm{{12}}$\\
  \hline
  Co$_\mathrm{{1}}$& 1  0  0& 0  1  0& 0  0  1& 0  0  0& 0  0  0& 0  0  0& 0  0  0& 0  0  0& 0  0  0& 1  0  0& 0  1  0& 0  0  1\\
  Co$_\mathrm{{2}}$& 1  0  0& 0  1  0& 0  0 -1& 0  0  0& 0  0  0& 0  0  0& 0  0  0& 0  0  0& 0  0  0& 1  0  0& 0  1  0& 0  0 -1\\
  Co$_\mathrm{{3}}$& 0  1  0&-1  0  0& 0  0  1& 0  0  0& 0  0  0& 0  0  0& 0  0  0& 0  0  0& 0  0  0& 0 -1  0& 1  0  0& 0  0 -1\\
  Co$_\mathrm{{4}}$& 0  1  0&-1  0  0& 0  0 -1& 0  0  0& 0  0  0& 0  0  0& 0  0  0& 0  0  0& 0  0  0& 0 -1  0& 1  0  0& 0  0  1\\

  Co$_\mathrm{{5}}$& 0  0  0& 0  0  0& 0  0  0&-1  0  0& 0  1  0& 0  0  1&-1  0  0& 0  1  0& 0  0  1& 0  0  0& 0  0  0& 0  0  0\\
  Co$_\mathrm{{6}}$& 0  0  0& 0  0  0& 0  0  0&-1  0  0& 0  1  0& 0  0 -1&-1  0  0& 0  1  0& 0  0 -1& 0  0  0& 0  0  0& 0  0  0\\
  Co$_\mathrm{{7}}$& 0  0  0& 0  0  0& 0  0  0& 0  1  0& 1  0  0& 0  0  1& 0 -1  0&-1  0  0& 0  0 -1& 0  0  0& 0  0  0& 0  0  0\\
  Co$_\mathrm{{8}}$& 0  0  0& 0  0  0& 0  0  0& 0  1  0& 1  0  0& 0  0 -1& 0 -1  0&-1  0  0& 0  0  1& 0  0  0& 0  0  0& 0  0  0\\
  \end{tabular}
  \end{ruledtabular}
\end{table}
\begin{figure}[H]
	\centering
	\includegraphics[width=0.96\textwidth]{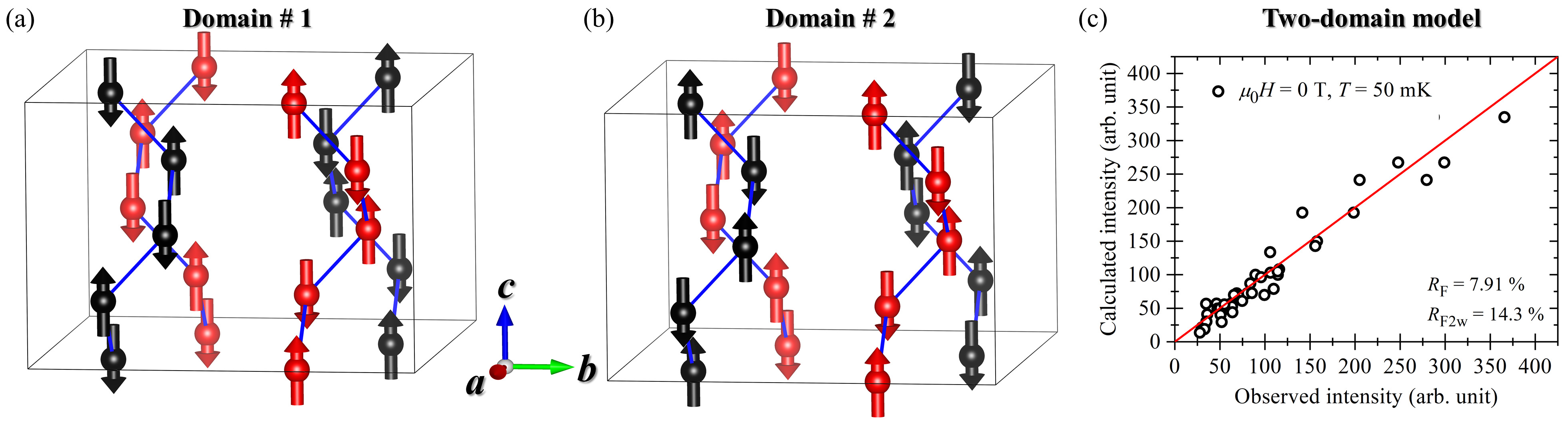}
	\caption{(a-b) Magnetic structure of SrCo$_{2}$V$_{2}$O$_{8}$ in the two magnetic domains of the N\'eel phase. The screw chains formed by Co$_\mathrm{{n}}$ ($n$ = 1-4) and Co$_\mathrm{{n}}$ ($n$ = 5-8) sites are marked in red and black, respectively. (c) Magnetic structure refinement in the N\'eel phase at $T$ = 50 mK and $\mu_\mathrm{0}H$ = 0 T. The R-factors ($R_\mathrm{F}$, $R_\mathrm{F2w}$) which characterise the quality of our refinement are also listed.}	
	\label{fig:ref}
\end{figure}

We have collected 46 magnetic reflections for the magnetic structure determination at 50 mK and 0 T. We first discuss the two-domain solution proposed in Ref.~\onlinecite{Canevet}; the corresponding magnetic structures are illustrated in Fig. \ref{fig:ref}a\ $\&$\ b. This scenario is allowed because the basis functions of Co$_\mathrm{{n}}$ ($n$ = 1 - 4) are independent of those of Co$_\mathrm{{n}}$ ($n$ = 5 - 8) (Table\ \ref{tab:I}). Following this approach, we constrain all the Co sites to have an identical amplitude of magnetic moment, while their orientations are decided by symmetry \cite{Canevet}. Initially, we included all the basis functions of $\Gamma_\mathrm{{5}}$ in the refinement. The coefficients for the ones contributing to a finite moment in the $ab$ plane (highlighted in grey in Table\ \ref{tab:I}) were found to be too weak to be resolved from our data, but we cannot exclude their existence.. This agrees well with the Ising-like anisotropy revealed in the magnetization and neutron powder diffraction studies \cite{He2,Bera}. Moreover, we could not detect any contribution from $\Psi_\mathrm{{6}}$ and $\Psi_\mathrm{{12}}$ within the resolution. As a result, only $\Psi_\mathrm{{3}}$ and $\Psi_\mathrm{{9}}$ were adopted in our final refinement; this fits the description of the N\'eel order predicted by Eq. \ref{Ham}, wherein the spins are antiferromagnetically coupled along the chain (Fig. \ref{fig:ref}a\ $\&$\ b). As shown in Fig. \ref{fig:ref}c, the two-domain solution reproduces the experimental observations well. The domain populations in our sample are 42(3) $\%$ and 58(3) $\%$ for Domain $\#$ 1 and Domain $\#$ 2 (Fig. \ref{fig:ref}a\ $\&$\ b), respectively. The refined moment along the $c$ axis is 1.81(4) $\mu_{\mathrm{B}}$ per Co. We note that this value is lower than the 2.1 - 2.3 $\mu_{\mathrm{B}}$ per Co at 1.5 K reported in the powder study \cite{Bera}, which may be due to variation in the sample quality.

\begin{figure}[b]
	\centering
	\includegraphics[width=0.75\textwidth]{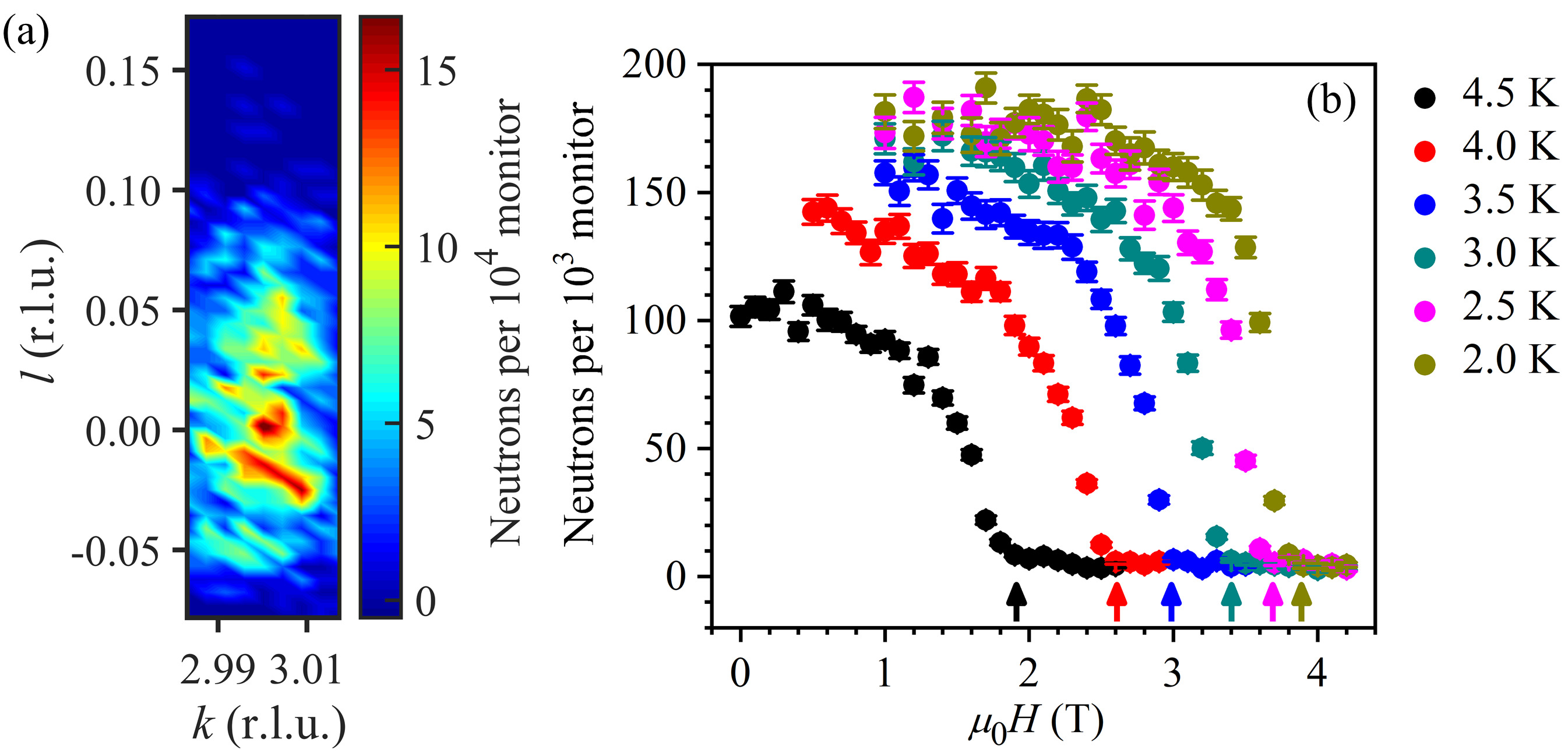}
	\caption{(a) The diffraction pattern of the (2, 3, 0) reflection recorded using a PSD\ \cite{rita} at 4.0 K and 0.5 T. The streak-like nature is an artifact of the vertically focusing monochromator. (b) The summed intensity of (2, 3, 0) between 2.0\,K and 4.5\,K as a function of magnetic field.}	
	\label{fig:3}
\end{figure}

Our single crystal data allows us to test a single-domain solution in which the aforementioned moment constraint for the Co sites is released\ \cite{Canevet}. We obtained 2.6(1) and -0.20(9) $\mu_{\mathrm{B}}$ for the Co$_\mathrm{{n}}$ ($n$ = 1 - 4) and Co$_\mathrm{{n}}$ ($n$ = 5 - 8) sites (Table \ref{tab:I}), respectively. Since the single-domain refinement also reproduces the experimental observations well ($R_\mathrm{F}$ = 7.85 $\%$, $R_\mathrm{F2w}$ = 14.6 $\%$), we cannot rule out this possibility. Local probes, such as nuclear magnetic resonance\ \cite{Kawasaki2}, are needed to further clarify the magnetic structure in this compound in the future.


While applying a magnetic field (\textit{H}\ $\parallel$\ \textit{c} axis), magnetization and heat capacity measurements suggest that SrCo$_{2}$V$_{2}$O$_{8}$ undergoes a field-induced order-disorder transition between $T_\mathrm{{N}}$ and 2.0\,K\ \cite{He2}. We studied the magnetic field dependence of the (2, 3, 0) reflection using the PSD on RITA-II. A typical diffraction pattern measured at 4.0\,K and 0.5\,T is shown in Fig.\,\ref{fig:3}a. We then studied the magnetic field dependence of this reflection at several temperatures between 2.0 K and 4.5 K. These observations are summarized in Fig.\,\ref{fig:3}b. A field-induced order-disorder transition, which has been observed in the sister compound BaCo$_{2}$V$_{2}$O$_{8}$\ \cite{He3,Canevet,Kawasaki}, is also clearly present in SrCo$_{2}$V$_{2}$O$_{8}$. We note that we could not detect any field-induced change in the magnetic modulation vector at all temperatures and fields discussed in this section.

\section{Magnetic-field-induced phase transitions below 2.0 K}\label{LT}
\begin{figure}[H]
	\centering
	\includegraphics[width=0.85\textwidth]{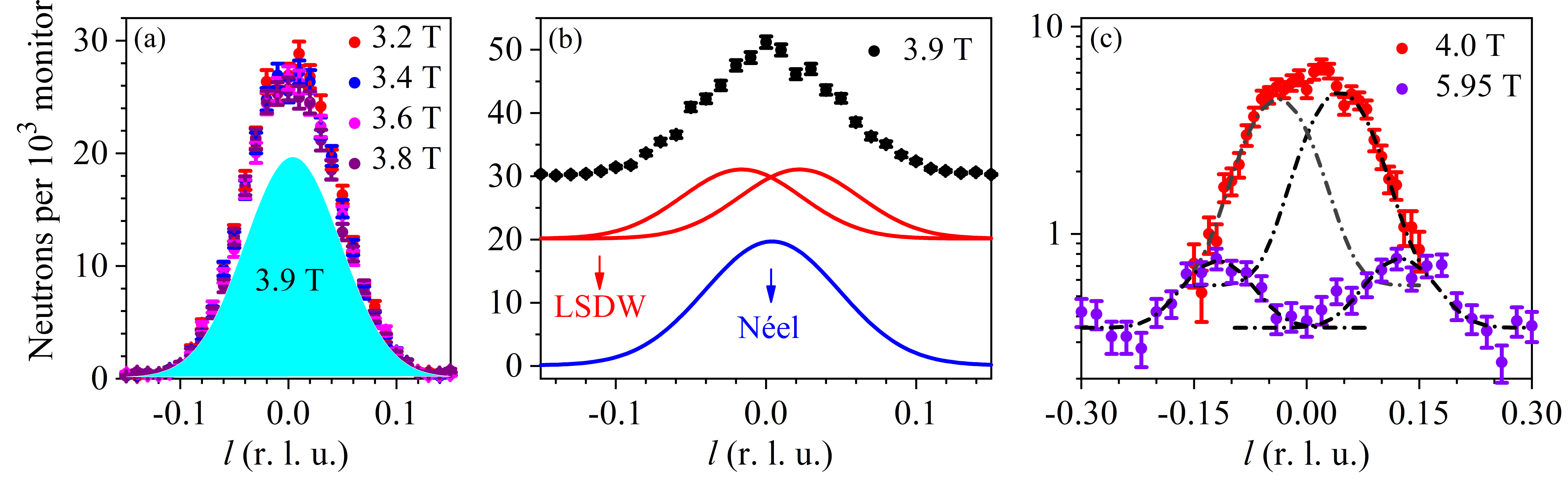}
	\caption{ Magnetic field dependence of the (2, 3, \textit{l}) reflection (\textit{l}-scan) at 75\,mK and (a) 3.2\,T\ $\leqslant$\ $\mu_\mathrm{{0}}H$\ $\leqslant$\ 3.8\,T, (b) $\mu_\mathrm{{0}}H$\ =\ 3.9\,T, and (c) $\mu_\mathrm{{0}}H$\ =\ 4.0 and 5.95\,T. The shaded area in (a) denotes the fit to the data at 3.9\,T for comparison. The curves displayed in (b) have been shifted vertically. The solid lines in (b) and (c) are numerical fits (see the main text). The counting time for each 10$^{3}$ neutron monitor is about 0.34 seconds.}	
	\label{fig:4}
\end{figure}
 
We now study the magnetic field response of the C-AFM order at temperatures down to 50\,mK. Figure\ \ref{fig:4} shows the (2, 3, \textit{l}) reflection at 75\,mK measured between 3.2\,T and 5.95\,T on TriCS. Below 3.9\,T, this reflection hardly changes and is centered at \textit{l}\ = 0 (Fig.\,\ref{fig:4}a).\ The average FWHM of the peaks in this region is 0.098\ r.\,l.\,u.; it is equal to the instrumental resolution within the errors (see Section\ \ref{HT}). At 3.9\,T, the intensity weakens (Fig.\,\ref{fig:4}a). At higher fields, we clearly see two peaks modulated by \textit{\textbf{k}}$_{IC}$ = (0, 0, 1\,$\pm$\,$\delta{l}$), featuring an IC-AFM phase (Fig.\,\ref{fig:4}c). While magnetic refinement is needed to properly determine the nature of this IC-AFM order, one plausible scenario, based on the splitting of the magnetic reflection along $\textbf{c}^{\star}$, is LSDW order, as demonstrated in the related compound BaCo$_{2}$V$_{2}$O$_{8}$\,\cite{Canevet,Kimura2}.

We have fitted the magnetic reflection at 3.9 T using three models (Fig.\,\ref{fig:4}b). In the first model, we apply a single Gaussian function centered at \textit{l}\ =\ 0 to this profile,\ meaning that the system is still in the C-AFM state. We obtain a broadened reflection with FWHM\ =\ 0.105(2)\,r.\,l.\,u.\ (Fig.\,\ref{fig:4}b). In the second model, we apply two Gaussian functions centered at \textit{l}\ =\ $\pm\delta{}l$. This model, which produces a FWHM of 0.09(1)\ r.\,l.\,u., corresponds to a single IC-AFM phase (Fig.\,\ref{fig:4}b). The third model, in which we assume the coexistence of the C-AFM and IC-AFM reflections, gives a FWHM of 0.07(1)\,r.\,l.\,u..\ This value is much smaller than the instrumental resolution 0.098(1)\,r.\,l.\,u.\ (Section\,\ref{HT}); we therefore rule out the third model. While the second model fits the observations in BaCo$_{2}$V$_{2}$O$_{8}$\ \cite{Kimura,Kimura2}, we cannot rule out the presence of a short-range ordered C-AFM phase at 3.9\,T within the experimental resolution.

We could only resolve the two IC reflections when $\mathrm{\mu_{0}}$\textit{H}\ $\geqslant$\ 4.0\,T, based on which we conclude that the IC-AFM state sets in around 3.9\,T in SrCo$_{2}$V$_{2}$O$_{8}$. The IC peak is resolution limited in an \textit{l}-scan at all magnetic fields measured, indicating long-range spin correlation along the \textit{c}-axis. This is consistent with the observations of the LSDW order in BaCo$_{2}$V$_{2}$O$_{8}$\,\cite{Kimura2, Canevet}. The IC reflection could be detected up to the highest field measured on TriCS (5.95\,T). We also tracked the temperature dependence of this phase.\ These results are summarized in the magnetic field versus temperature phase diagram displayed in Fig.\,\ref{fig:9} later in the paper.

\begin{figure}[t]
	\centering
	\includegraphics[width=0.95\textwidth]{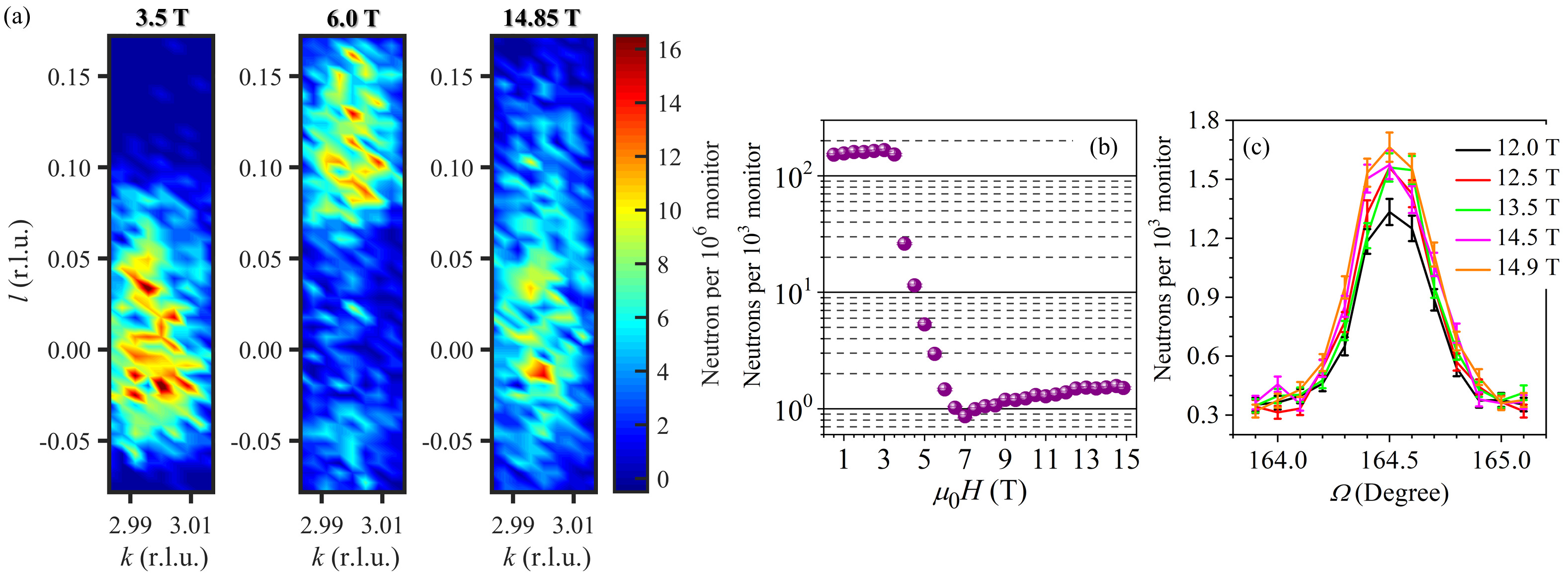}
	\caption{(a) The (2, 3, 0) reflection at selective magnetic fields. A scale factor of 1/100 has been applied to the data at 3.5 T.  (b) Magnetic field dependence of the summed intensity. (b) Rocking curves of the (2, 3, 0) reflection for $\mu_\mathrm{{0}}H$\ $\geqslant$\ 12.0\,T. All the data were collected at 120 mK.}
	\label{fig:6}
\end{figure}

To explore the evolution of the IC-AFM order above 5.95\,T in SrCo$_{2}$V$_{2}$O$_{8}$, we performed further measurements on RITA-II using a 15\,T vertical cryomagnet.\ Neither the cryomagnet nor detector could be tilted, meaning that the neutron scattering intensity from the IC-AFM structure will be weakened by the Lorentz factor. However, by taking advantage of the PSD, we were still able to resolve a partial IC diffraction spot, and thus track its evolution. 

Based on the data from TriCS, the diffraction spot of the (2, 3, 0) reflection in the PSD is expected to split vertically, i.e. along $\textbf{c}^\star$, as the magnetic field is driven across the C-AFM to IC-AFM phase boundary. At 6.0\ T, while the -$\delta{l}$ satellite peak goes out of the detection range of the PSD, the +$\delta{l}$ satellite peak can be clearly observed (Fig.\,\ref{fig:6}a). The split can be resolved up to 6.5\ T. Interestingly, the (2, 3, 0) reflection is recovered above 7.0\ T (Fig.\,\ref{fig:6}a). Due to the geometry limitation of the 15\,T cryomagnet, we could not check the scattering signal in an extensive reciprocal space region. As a result, a multi-$k$ modulation cannot be ruled out for the high field emergent phase. However, similar reentrant behaviour has been observed in BaCo$_{2}$V$_{2}$O$_{8}$ and was interpreted as a sign of the LSDW to transverse AFM order crossover\,\cite{Grenier2}. We plot the summed intensity versus magnetic field curve in Fig.\,\ref{fig:6}b, in which the drastic drop above 4.0\,T fits the C-AFM to IC-AFM order transition illustrated in Fig.\ \ref{fig:4}. We note that the rate of this drop is overestimated due to the fact that only the partial double peak profiles could be resolved above 5.0\,T (Fig.\,\ref{fig:6}a). The intensity reaches its minimum at 7.0\,T, after which it increases monotonously with the magnetic field until 12.5\,T. The reemergence of the scattering intensity above 7.0\,T (Fig.\,\ref{fig:6}b) is also consistent with the transverse AFM order observed in BaCo$_{2}$V$_{2}$O$_{8}$\ \cite{Grenier2}. At 7.0 T, we performed two additional measurements with a much longer counting time. We could resolve a signal at 100\,mK, which fades away at 750\,mK. However, this signal is too weak to check the coexistence of the IC-AFM and emergent AFM order reported in BaCo$_{2}$V$_{2}$O$_{8}$\ \cite{Grenier2}. The emergent AFM order in SrCo$_{2}$V$_{2}$O$_{8}$ is fully stabilized at 12.5\,T and above; the intensity remains unchanged up to the highest field probed (14.9\,T, see Fig.\,\ref{fig:6}c).


\section{Discussion}\label{Discussion}\label{Discussion}

\begin{figure}[H]
	\centering
	\includegraphics[width=0.9\textwidth]{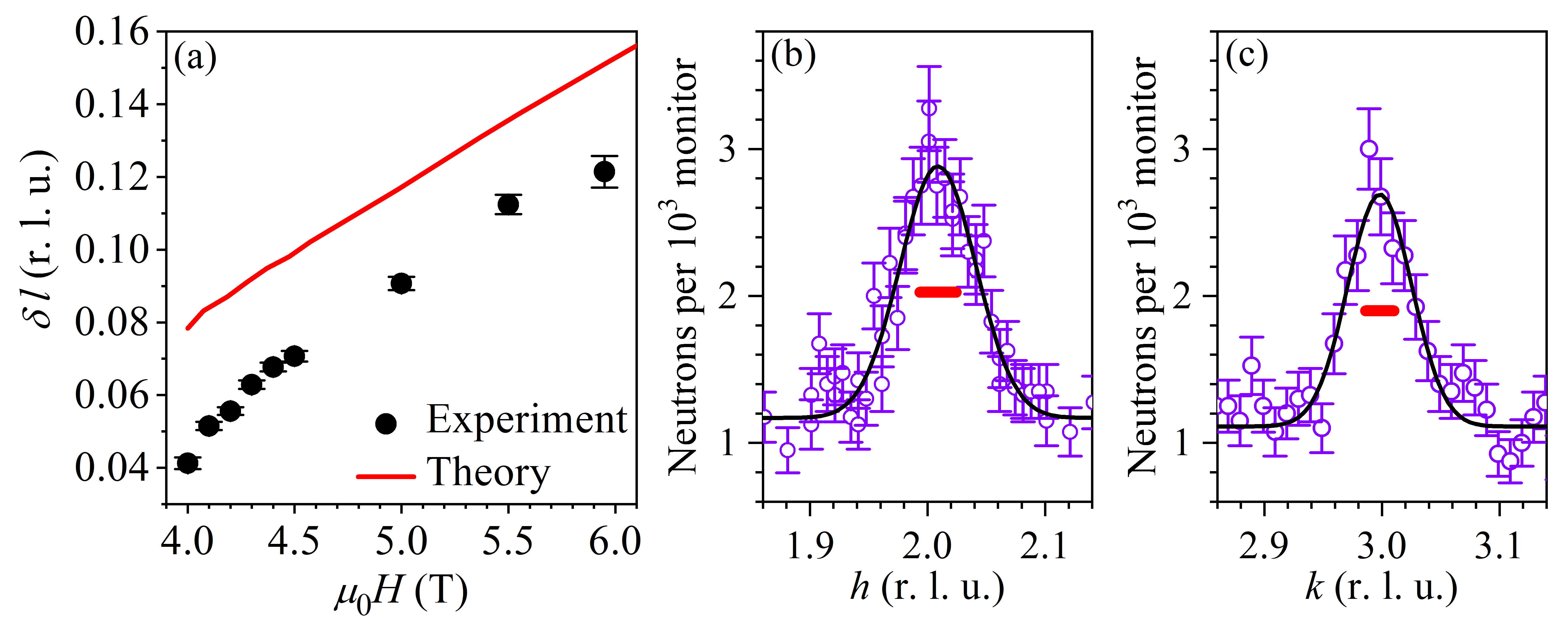}
	\caption{(a) Magnetic field dependence of the incommensurate propagation vector $\delta{l}$ at 75\,mK. The filled circles are the experimental observations, and the red line is  the theoretical values deduced from Eq.\,\ref{kz} and Ref.~\onlinecite{Okutani}. (b) \textit{h}- and (c) \textit{k}- scans of the (2, 3, $\delta{l}$) reflection at $\mu_\mathrm{{0}}H$\ =\ 5.0\,T and \textit{T}\ =\ 150\,mK. The black lines are Gaussian fits. The red bars are the instrumental resolution determined at zero field (see Section\,\ref{HT}).}	
	\label{fig:5}
\end{figure}

\begin{figure}
	\centering
	\includegraphics[width=0.9\textwidth]{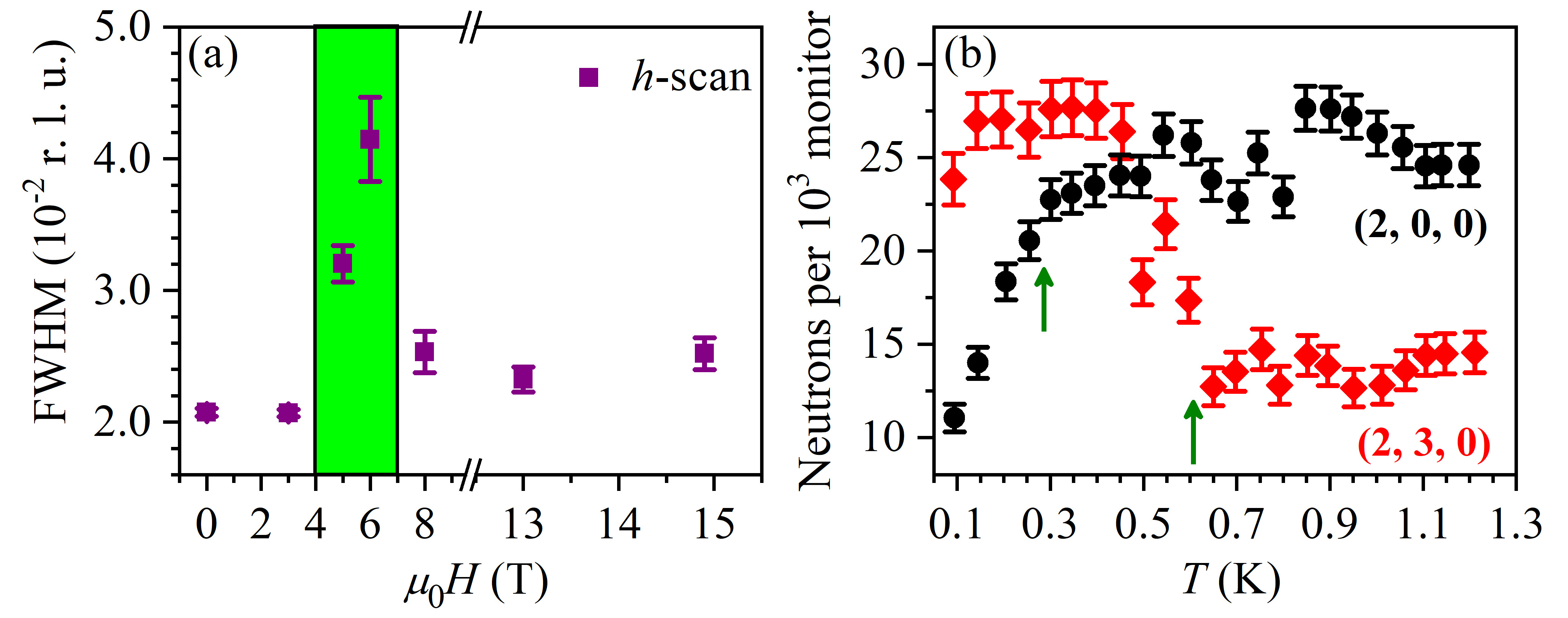}
	\caption{(a) Magnetic field dependence of the FWHM of the (2,\ 3,\ 0) reflection at 120\,mK measured in a \textit{h}-scan on RITA-II. The green shaded area marks the LSDW region. (b) Temperature dependence of the (2,\ 0,\ 0) and (2,\ 3,\ 0) reflections at 7.5\,T. A scale factor of 40 has been applied to the (2,\ 3,\ 0) reflection. The green arrows mark the transition temperatures described in the main text.}	
	\label{fig:8}
\end{figure}

In the IC-AFM state, increasing the magnetic field pulls the two IC peaks further apart (Fig.\,\ref{fig:4}c). It has been proposed that the IC-AFM order results from the condensation of the longitudinal spin fluctuation of a TLL, meaning that its incommensurability $\delta{l}$ should exactly follow Eq.\,\ref{kz}\,\cite{Kimura,Kimura2,Canevet,Okunishi}; this has been verified in BaCo$_{2}$V$_{2}$O$_{8}$\,\cite{Kimura2, Canevet}. To check this scenario in SrCo$_{2}$V$_{2}$O$_{8}$, we compare our fitted $\delta{l}$ at 75\,mK with the values predicted by Eq.\,\ref{kz} (Fig.\,\ref{fig:5}a). In this plot, the Land\'e \textit{g}-tensor along the $c$ axis ($g_\mathrm{c}$ = 6.1), Van Vleck paramagnetism correction factor (0.014\,$\mu_\mathrm{{B}}$/T) and longitudinal uniform magnetization $M_\mathrm{{z}}$ measured between 1.3 K and 1.9 K were used to produce the theoretical $\delta{l}$; these values have been reported in Ref.~\onlinecite{Okutani}. In sharp contrast to the good match between the experiment and theory in BaCo$_{2}$V$_{2}$O$_{8}$\ \cite{Kimura2,Canevet}, all the observed $\delta{}l$ in SrCo$_{2}$V$_{2}$O$_{8}$ are much lower than the predicted values (Fig.\,\ref{fig:5}a). This discrepancy casts some doubt on the applicability of the TLL theory in interpreting the IC-AFM order in this case. However, the low-energy fractional magnetic excitations in this compound, e.g. spinons, (anti)psinons and Bethe strings, have been proven to be well described by Eq.\,\ref{Ham} \cite{Bera3,Wang,Wang2}. Moreover, the N\'eel ordering temperature of SrCo$_{2}$V$_{2}$O$_{8}$ ($T_\mathrm{N}$ = 5.0 K) is lower than that of BaCo$_{2}$V$_{2}$O$_{8}$ ($T_\mathrm{N}$ = 5.5 K)\,\cite{Canevet}, supporting the stronger 1D character in the former. Based on these facts, it is less likely that the TLL theory fails in SrCo$_{2}$V$_{2}$O$_{8}$. We therefore propose another possible explanation here. The magnetization measurements for obtaining $g_{c}$ and $M_\mathrm{{z}}$ were performed between 1.3\,K and 1.9\,K\ \cite{Okutani}. To study $M_\mathrm{{z}}$ at lower temperatures, we measured the nuclear reflections (2,\ 0,\ 0) and (4,\ 0,\ 0). At high fields, additional neutron counts can be resolved on top of (2,\ 0,\ 0), while (4,\ 0,\ 0) is not affected up to 14.9\,T (not shown here). This indicates the ferromagnetic origin of the weak field-induced intensity at (2,\ 0,\ 0)\ \cite{Canevet,Grenier2}; the non-resolvable change at (4,\ 0,\ 0) is presumably due to the reduction of the magnetic form factor. Interestingly, our measurements on the (2,\ 0,\ 0) reflection at 7.5 T clearly reveal an intensity drop below $\sim$ 0.3\,K (Fig.\,\ref{fig:8}a). This suppression is independent of the emergent AFM order, as the latter sets in at 0.6\,K. Although we did not measure the temperature dependence of this reflection at lower fields due to the much weaker signal (0.1-0.2\,$\mu_\mathrm{{B}}$/Co\,\cite{Okutani}), we believe that the suppressed $M_\mathrm{{z}}$ might persist in the IC-AFM region, and be responsible for the discrepancy shown in Fig.\,\ref{fig:5}a. 

We investigated the transverse spin correlation as a function of magnetic field (Fig.\,\ref{fig:5}b\,\&\,c measured on TriCS and Fig.\,\ref{fig:8}b measured on RITA-II). Unlike the three-dimensional long-range LSDW order in BaCo$_{2}$V$_{2}$O$_{8}$\ \cite{Kimura2}, the (2, 3, $\pm\delta{}l$) peak is not resolution limited along both the \textit{h}- and \textit{k}- directions in SrCo$_{2}$V$_{2}$O$_{8}$, which suggests short-range spin correlation in the \textit{ab} plane. Interestingly, the transverse long-range spin correlation is recovered in the emergent AFM state (Fig.\,\ref{fig:8}b). The suppressed longitudinal uniform magnetization $M_\mathrm{{z}}$ and transverse short-range spin correlation indicate that the IC-AFM order in SrCo$_{2}$V$_{2}$O$_{8}$ is distinct from the perfect LSDW order observed in BaCo$_{2}$V$_{2}$O$_{8}$.

In a TLL with Ising-like anisotropy, the critical magnetic field ($H_\mathrm{{c}}$) at which the crossover between the longitudinal and transverse spin fluctuations occurs scales linearly with the intrachain exchange strength (\textit{J} in Eq.\ \ref{Ham})\ \cite{Okunishi}. The LSDW order in an Ising-like quasi 1D quantum magnet results from the condensation of the longitudinal mode while the interactions between the chains become energetically relevant. An important conclusion revealed by the mean-field theory is that the collapse of the LSDW order does not coincide with $H_\mathrm{{c}}$, but shifts to a higher value\ \cite{Okunishi}. Therefore, the LSDW order is expected to be more robust in a system with a larger \textit{J}. Based on the high field magnetization and inelastic neutron scattering investigations, \textit{J} is larger in SrCo$_{2}$V$_{2}$O$_{8}$ than that in BaCo$_{2}$V$_{2}$O$_{8}$\ \cite{Kimura3, Okutani, Grenier, Bera,Matsuda}. However, the IC-AFM order in SrCo$_{2}$V$_{2}$O$_{8}$ turns out to be more fragile (Fig.\,\ref{fig:6}b). We note that the experimental critical field for the LSDW to transverse AFM order transition in BaCo$_{2}$V$_{2}$O$_{8}$\ \cite{Grenier2} already deviates from the mean-field prediction in Ref.~\onlinecite{Okunishi}. The even more significant deviation in SrCo$_{2}$V$_{2}$O$_{8}$ revealed in our study further stresses the inadequacy of the interchain mean-field theory in addressing the complex magnetism in these systems. 

\section{Summary}
 
\begin{figure}[H]
	\centering
	\includegraphics[width=0.6\textwidth]{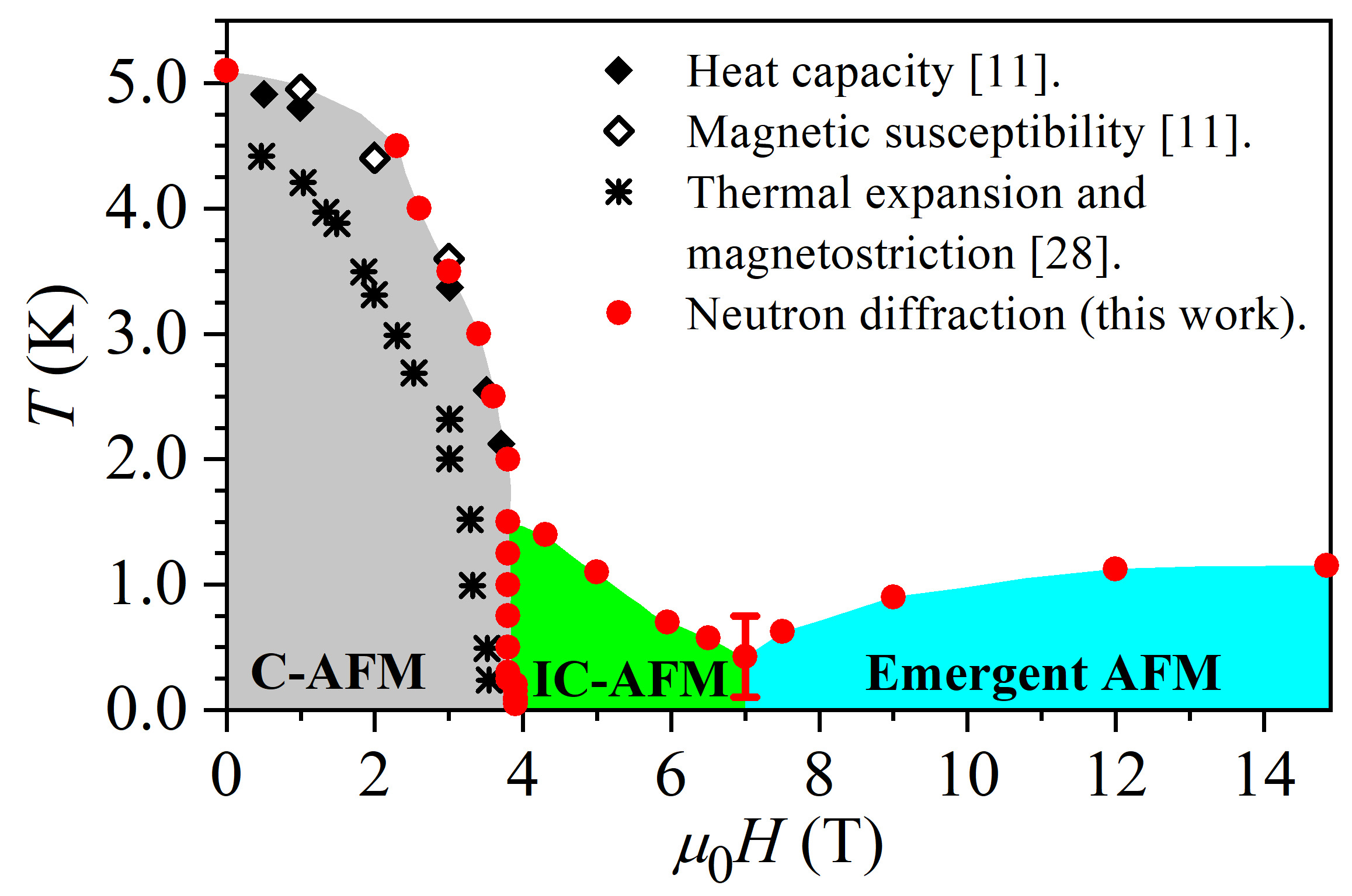}
	\caption{Magnetic field versus temperature phase diagram of SrCo$_{2}$V$_{2}$O$_{8}$.}	
	\label{fig:9}
\end{figure}

In conclusion, we have employed single-crystal neutron diffraction to map out the magnetic phase diagram of SrCo$_{2}$V$_{2}$O$_{8}$ up to 14.9\,T and down to 50\,mK. As shown in Fig.\,\ref{fig:9}, our results are in excellent agreement with the previous magnetization and heat capacity investigation on the same crystal \cite{He2}. The deviation from the thermal expansion and magnetostriction study, in which the single-crystal was grown by a different method\,\cite{Niesen2}, could come from the variation in sample quality. This system is composed of weakly coupled \textit{S}\ =\ 1/2 XXZ spin chains with Ising-like anisotropy. Like its related counterpart BaCo$_{2}$V$_{2}$O$_{8}$, SrCo$_{2}$V$_{2}$O$_{8}$ shows multiple magnetic-field-driven phase transitions that reflect the quantum critical nature of the spins in a TLL state. In addition to the similarity, we have identified several unique signatures for the spin states in SrCo$_{2}$V$_{2}$O$_{8}$, including the fragility of IC-AFM order, loss of three-dimensional long-range spin correlation in the IC-AFM region and suppression of uniform magnetization along the \textit{c} axis at low temperatures.

Our observations highlight the complex magnetic properties in these systems and evidence the inadequacy of the interchain mean-field theory. Further investigations, e.g. magnetic refinement, density functional theory calculations and inelastic neutron spectroscopy measurements, are in high demand to shed more lights on the magnetic structures in the C-AFM, IC-AFM, and emergent AFM states, as well as the interchain couplings in SrCo$_{2}$V$_{2}$O$_{8}$.

%



\begin{acknowledgments}
We acknowledge the UK EPSRC for funding under grant number EP/J016977/1. This work is based on experiments performed at the Swiss spallation neutron source SINQ, Paul Scherrer Institute, Villigen, Switzerland. One of the authors, Z. He thanks the National Natural Science Foundation of China (No. 21573235) and the Chinese Academy of Sciences under Grant No.KJZD-EW-M05 for financial support.
\end{acknowledgments}

\bibliography{NJP}
\end{document}